\documentclass[%
aip,
rsi,%
amsmath,amssymb,
 reprint,%
]{revtex4-1}

\usepackage{graphicx}
\usepackage{dcolumn}
\usepackage{bm}
\usepackage[mathlines]{lineno}
\usepackage{color}
\begin{document}

\preprint{AIP/123-QED}

\title{High-pressure cell for simultaneous dielectric and neutron spectroscopy}

\author{Alejandro Sanz}\email{asanz@ruc.dk}
\author{Henriette Wase Hansen}
\author{Bo Jakobsen}
\author{Ib H. Pedersen}
\affiliation{Glass and Time, IMFUFA, Department of Science and Environment, Roskilde University, Postbox 260, DK-4000 Roskilde, Denmark}
\author{Simone Capaccioli}
\affiliation{Department of Physics, University of Pisa, 56127 Pisa, Italy}
\author{Karolina Adrjanowicz}
\author{Marian Paluch}
\affiliation{Institute of Physics, University of Silesia, ul. Uniwersytecka 4, 40-007 Katowice, Poland}
\author{Julien Gonthier}
\author{Bernhard Frick}
\author{Eddy Leli\`evre-Berna}
\affiliation{Institut Laue-Langevin, CS 20156, 38042 Grenoble Cedex 9, France}
\author{Judith Peters}
\affiliation{Institut Laue-Langevin, CS 20156, 38042 Grenoble Cedex 9, France}
\affiliation{Universit\'e Grenoble Alpes, CNRS, LIPhy, 38000 Grenoble, France}
\author{Kristine Niss}
\affiliation{Glass and Time, IMFUFA, Department of Science and Environment, Roskilde University, Postbox 260, DK-4000 Roskilde, Denmark}

\date{\today}

\begin{abstract}

In this article we report on the design, manufacture and testing of a high-pressure cell for doing simultaneous dielectric and neutron spectroscopy. This cell is a unique tool for studying dynamics on different timescales, from kilo- to picoseconds, covering universal features such as the $\alpha$ relaxation and fast vibrations at the same time. The cell, constructed in cylindrical geometry, is made in high-strength aluminum alloy and operates up to 500~MPa in a temperature range between roughly 2 and 320 K. In order to measure the scattered neutron intensity and the sample capacitance simultaneously, a cylindrical capacitor is positioned within the bore of the high-pressure container. The capacitor consists of two concentric electrodes separated by insulating spacers. The performance of this setup has been successfully verified by collecting simultaneous dielectric and neutron spectroscopy data on dipropylene glycol, using both backscattering and time-of-flight instruments. We have carried out the experiments at different combinations of temperature and pressure in both supercooled liquid and glassy state.

\end{abstract}


\maketitle


\section{Introduction}

Supercooled liquids are metastable with respect to the crystalline state and induced e.g. by rapid cooling below their melting point $T_{m}$ to avoid nucleation and crystal growth. Alternatively, compression can be utilized to create metastable liquids below $P_m$. If temperature keeps decreasing (or pressure increases), the structural relaxation is found to be more and more hindered and eventually the rearrangement of the molecules becomes so slow that the equilibrium volume and enthalpy cannot be reached on experimental timescales. This marks the onset of the liquid-to-glass transition.\cite{Ediger,Berthier2} Pressure is an important variable for disentangling temperature and volumetric effects on the slowing down of the dynamics when approaching the glass transition. \cite{Roland} In the proximity of the glass transition temperature, $T_{g}$, the typical timescale for molecular rearrangements is minutes or even hours. Along with this extremely slow dynamics, glass-forming materials present other universal features such as fast vibrational processes as well as fast relaxations. It is now generally agreed that an appropriate comprehension of the glass transition must encompass fast and slow dynamics.\cite{Betancourt,Henriette} In order to study the broad range of dynamics covering molecular movements from picosecond to kilosecond timescales it is necessary to combine complementary techniques, such as dielectric and neutron spectroscopy.
 
For a better understanding of how slow and fast dynamics are correlated with each other at varying thermodynamic conditions, we present here a novel experimental setup to perform simultaneous dielectric and neutron spectroscopy. The cell, developed in collaboration between Roskilde University and Institut Laue-Langevin (ILL), has been designed for a temperature range of approximately 2-320~K and for a maximum pressure of 500~MPa, which is 5000 times atmospheric pressure. This setup is a unique tool for studying a variety of systems with dynamics on a large range of timescales, such as viscous liquids, polymers and proteins. By performing simultaneous dielectric and neutron spectroscopy experiments, we are able to monitor slow and fast dynamics under exactly the same conditions, precluding unwanted external effects associated with different environments, as for example shape and material of the container, and heat and pressure transfer. The simultaneous measurements are especially important when the sample is not stable; for instance due to physical or chemical processes such as aging, crystallization, chemical degradation or water uptake.

In dielectric spectroscopy (DS), the sample is placed between two conducting electrodes, creating a capacitor. This technique measures the complex capacitance $C^*$ as a function of the applied frequency $\nu$, typically from mHz to MHz. From the capacitance, the complex dielectric permittivity of the sample $\varepsilon^*(\nu)=\varepsilon'(\nu)+i\varepsilon''(\nu)$ can be calculated, either through the capacitance of the empty capacitor or from knowing the geometry of the capacitor. When the material undergoes a relaxation process in response to an alternating electric field, the imaginary part of the dielectric permittivity, $\varepsilon''(\nu)$, referred to as the dielectric loss, shows a maximum as a function of frequency.\cite{Kremer} The position of this peak $\nu_\mathrm{max}$ provides information about the characteristic relaxation time $\tau$ of the underlying dynamics through the relation $\tau=(2\pi \nu_\mathrm{max})^{-1}$.
In incoherent neutron spectroscopy (NS) for samples with a large content of hydrogen, such as organic liquids, the incoherent dynamic structure factor, $S(\mathbf{Q},\omega)$ is measured as a function of energy transfer, $\hbar\omega$, and the momentum transfer, $\mathbf{Q}$, and provides information on the self-motion of atoms. Inelastic excitations in incoherent neutron scattering for hydrogenous samples are often seen as a broadening of the elastic peak commonly referred to as quasielastic neutron scattering (QENS).\cite{Bee} The energy resolution of the neutron spectrometer determines the timescale of the experiment. Spatial information of the dynamics can be obtained through the dependence of the dynamic structure factor on the momentum transfer $\mathbf{Q}=(\mathbf{k'}-\mathbf{k})$, where $\mathbf{k}$ is the wavevector of the incident neutron and $\mathbf{k'}$ is the scattered wavevector.\cite{Lovesey} The momentum transfer $\mathbf{Q}$ can be related to the scattering angle via $\mathbf{Q}=(4 \pi /\lambda)\sin(\theta/2)$, where $\lambda$ is the neutron wavelength and $\theta$ the scattering angle. Neutron backscattering spectroscopy (BS) normally covers molecular motions between $10^{-9}$ and $10^{-10}$~s, while instruments based on time-of-flight methods (TOF) provide information at the sub-nanosecond regime.\cite{Lovesey,Springer} Features such as vibrations, fast relaxations and the Boson peak have typically been detected by TOF spectrometers.

An advantage of combining dielectric spectroscopy and incoherent neutron scattering is that the optimal cell geometry is identical for both methods. The area should be as large as possible and the thickness of the sample should be small: For a capacitor with cylindrical geometry, the dielectric signal scales with the height of the capacitor and is inversely proportional to the distance between the electrodes. For incoherent QENS, the sample thickness must remain small in order to minimize multiple scattering effects,\cite{Bee} while a large sample area enhances the signal.

With this new setup, the sub-nanosecond dynamics can be measured simultaneously with the slow $\alpha$ relaxation seen in dielectric spectra, a universal fingerprint of the glass transition.\cite{Debenedetti,Colmenero} Given that the $\alpha$ relaxation stretches over several decades, it is still visible with DS when it enters into the NS window, at least as a high frequency tail.

In this paper we report on the design, manufacture and testing of a high-pressure (HP) cell for carrying out dielectric and neutron spectroscopy simultaneously. This setup can be utilized in different neutron spectrometers in order to combine DS and NS at different energy transfers and energy resolutions, thus accessing different timescales, in broad temperature and pressure ranges. We present the simultaneous collection of DS and NS data performed on the spectrometers IN16B and IN6 (ILL) for different combinations of temperature and pressure, demonstrating the utility of the technique.    

\section{Description of the high-pressure cell for simultaneous dielectric and neutron spectroscopy measurements}

An important requirement for constructing the cell is to reduce as much as possible the neutron attenuation. Materials with high tensile properties, low neutron absorption and incoherent scattering cross section are required.\cite{Klotz} We carried out finite element calculations and measured the transmission and background produced by different prototypes on the spectrometers IN6 \cite{IN6}, IN16B \cite{Frick,IN16b} and IN13 \cite{Natali} (ILL), in order to optimize the dimension, shape and type of material. Based on these results, the cell was built in cylindrical geometry using the aluminum alloy EN AW 7049A-T6. At 300 K, Al-7049A-T6 shows a yield stress of 0.53 GPa \cite{Klotz}. Given its high content of Al ($\sim$ mass 90 $\%$), Al-7049A-T6 shows a low incoherent neutron scattering cross section and thus a low background in the data. 

The development of high-pressure equipment in neutron facilities has been the subject of an intense research activity over the last decades (see, e.g., \cite{Klotz,Frick2002,Kuhs,Wang,Yang}).  
The design of our device was inspired by the high-pressure cell recently developed by Peters and co-workers, in collaboration with the Service for Advanced Neutron Environment (SANE) at the ILL.\cite{Peters} Proposed design utilized a piston or floating barrier to separate the sample volume from the pressure-transmitting liquid. They also discuss the special requirements that a sample holder must fulfill for an optimal performance. For example, the sample stick, which was also designed at the ILL, is not only responsible for holding the sample, but also for transmitting pressure onto the sample through a capillary that links the sample cell to a compressor outside the cryostat. The capillary must be thermally isolated as it passes the cold-point in the cryostat in order to avoid unwanted freezing of the pressure-liquid.\cite{Eddy}        

Our HP neutron cell is designed in a similar way, however, it integrates an additional component into the neutron beam; a cylindrical capacitor for measuring the dielectric properties of the sample. Within the bore of the HP container there is a coaxial cylinder surrounding a rod. Both the hollow cylinder and the rod, the outer and inner electrodes of the capacitor respectively, are also made of aluminum alloy Al-7049A-T6. This is challenging compared to ordinary HP cells for neutron scattering, since the electrodes have to be connected to the electronics through coaxial cables (RG178 with outer diameter 1.8 mm) which will be fed into the pressurized zone by a sealed plug with feedthroughs.

\subsection{Main components of the high-pressure cell}

\begin{figure*}
	\centering
	\includegraphics[trim = 15mm 50mm 15mm 20mm, clip, width=0.95\linewidth]{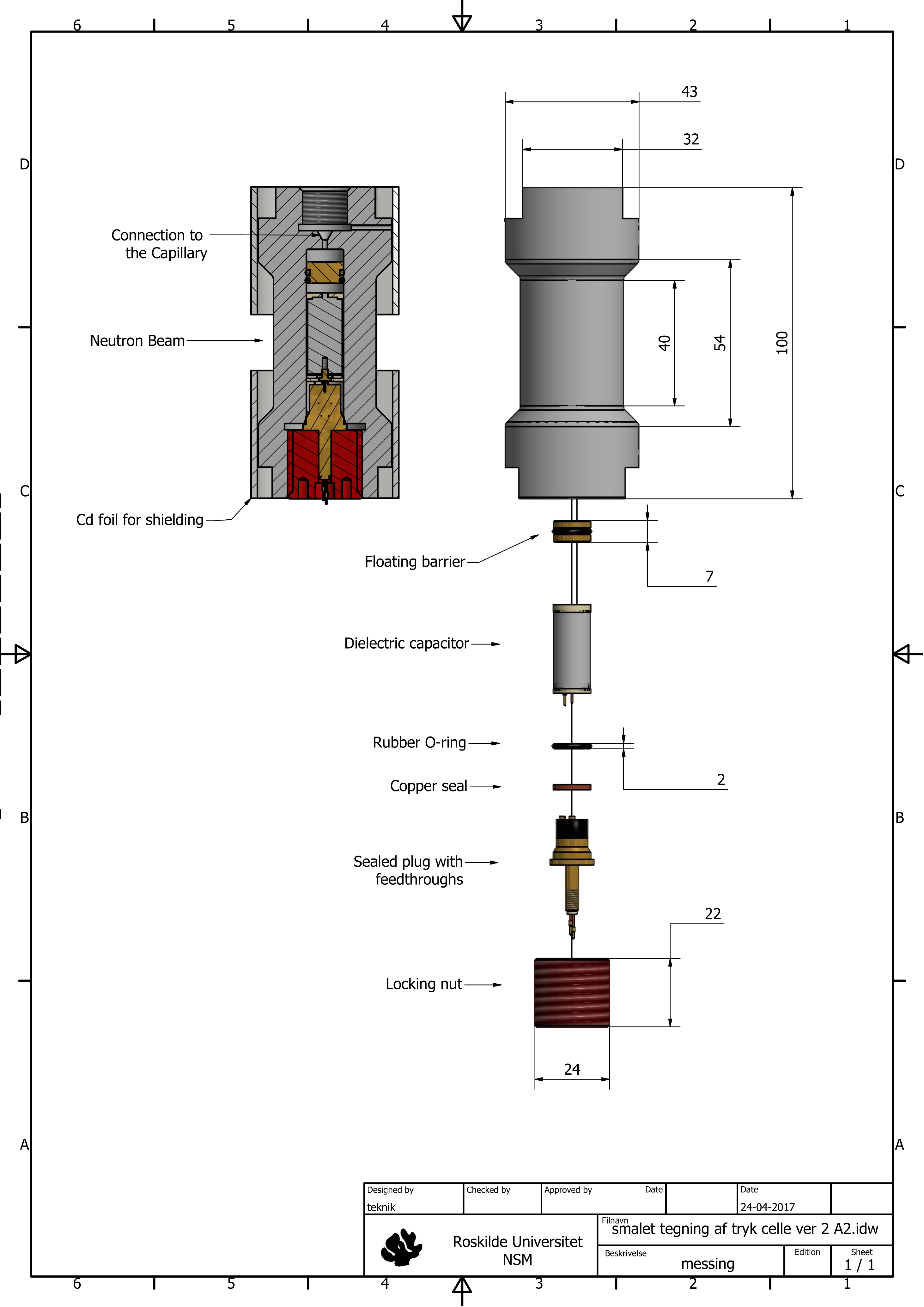}
	\caption{Exploded technical drawing of the high-pressure cell for simultaneous dielectric and neutron spectroscopy measurements. On the left-top corner, the figure displays a cross-sectional view of the assembled cell. Cadmium foil for shielding the wires, plug, floating barrier and the thickest parts of the cell are also illustrated. Dimensions are shown in mm.}
	\label{fig:Fig1}
	
\end{figure*}

\begin{figure}
	
	\includegraphics[trim = 2mm 0mm 0mm 5mm, clip, width=1.0\linewidth]{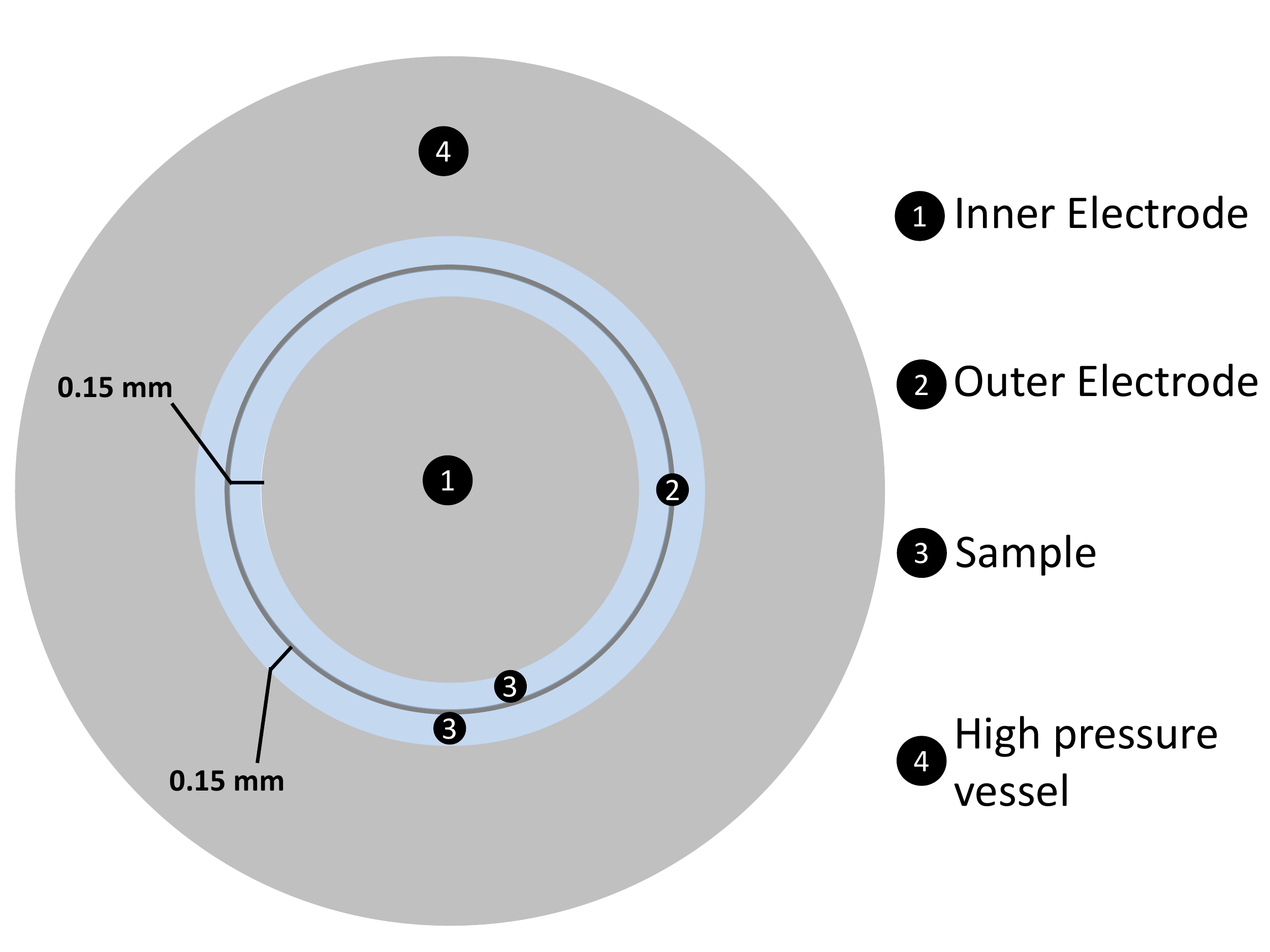}
	\caption{Schematic cross-sectional top view of the sample geometry for the high-pressure DS-NS cell. Dimensions are not to scale for the sake of clarity.}
	\label{fig:Fig4}
\end{figure}
     
The high-pressure DS-NS cell is presented in an exploded-view drawing (Fig.~\ref{fig:Fig1}). A hollow cylindrical monobloc with an inner diameter of 12 mm is the main-body piece. The outer diameter is reduced as much as possible in front of the beam section to minimize the neutron attenuation and diffuse scattering (background). It is 32~mm at the center, which is in the neutron beam, and 43~mm at the thickest parts. The top end is connected to a capillary injecting the pressure-transmitting liquid. In order to preclude unwanted mixing of the sample with the pressure-transmitting liquid, we use a floating barrier that consists of a disc made of brass surrounded at its middle part by a rubber O-ring seal. A reservoir of sample volume separates the floating barrier from the dielectric capacitor. In the other end of the sample cell, copper wires are fed into the pressurized zone through a plug made from brass sealed with an epoxy encapsulant and connected to the capacitor. Optimum tightness of the plug is achieved by the action of two seals, one made of copper and the other one made of rubber for high and low pressure ranges, respectively. The plug is held by a 22~mm long locking nut made of a copper-beryllium alloy. A cross-sectional view displayed in the left-top corner of Fig.~\ref{fig:Fig1} shows the location of cadmium foil for shielding the wires, sealed plug, floating barrier and the thickest parts of the HP container.

Figure~\ref{fig:Fig4} shows a schematic representation of the sample geometry illustrating the location of electrode-electrode and cell-electrode sample volumes. We expect identical behavior for the sample occupying these two compartments given the same thickness and that the high-pressure container and the electrodes are both made of the same aluminum alloy. For neutrons this results in a hollow cylinder sample geometry with a wall thickness of 0.3~mm.

\subsection{Strengthening of the high-pressure cell}

It is worth noting that compression may induce elastic and plastic deformation of the cell. The cell must operate in the elastic regime for optimal performance. In order to increase the maximum pressure at elastic regime, the HP cell was \textit{autofretted} by initial compression up to 530~MPa. The \textit{autofrettage} process is performed by applying pressure above the elastic region. For a monobloc cylinder, according to the Lam\'{e} equations and the distortion energy theory for plasticity criterion \cite{Klotz}, the bore of the cylinder will start to deform plastically at a pressure $P$:

\begin{equation}
	P=\sigma_{Y}\frac{\left(K^{2}-1\right)}{\sqrt{3}K^{2}}
	\label{eq:Eq1} 
\end{equation}

where $\sigma_{Y}$ is the yield stress of the material and $K = r_{o}/r_{i}$ is the wall ratio ($r_{o}$ and $r_{i}$ are the outer and inner radius of the cylinder respectively). By means of this strengthening process, it is possible to increase the maximum pressure at which the cell yields. In this procedure, the monobloc vessel is firstly bored with a slightly narrower inner diameter. Then it is pressurized above the limit defined by Eq.~\ref{eq:Eq1} in such a way that the material at the bore deforms plastically. Once the pressure is released, the material that was just deformed elastically will radially compress the area close to the bore that initially suffered plastic deformation. This raises the pressure limit at which the cell starts to yield. For a monobloc with $K > $ 2.2, the pressure limit is given by:

\begin{equation}
	P_{l}=2P
	\label{eq:Eq2} 
\end{equation}

Eq.~\ref{eq:Eq2} also determines the pressure for the \textit{autofrettage} process. Finally, the bore is machined to achieve the desired diameter and the cell will operate in an elastic regime as long as the pressure limit \textit{$P_{l}$} is not exceeded. 

\subsection{Dielectric capacitor and plug with electrical feedtroughs}

\begin{figure}
	
	\includegraphics[trim = 63mm 85mm 15mm 35mm, clip, width=1.15\linewidth]{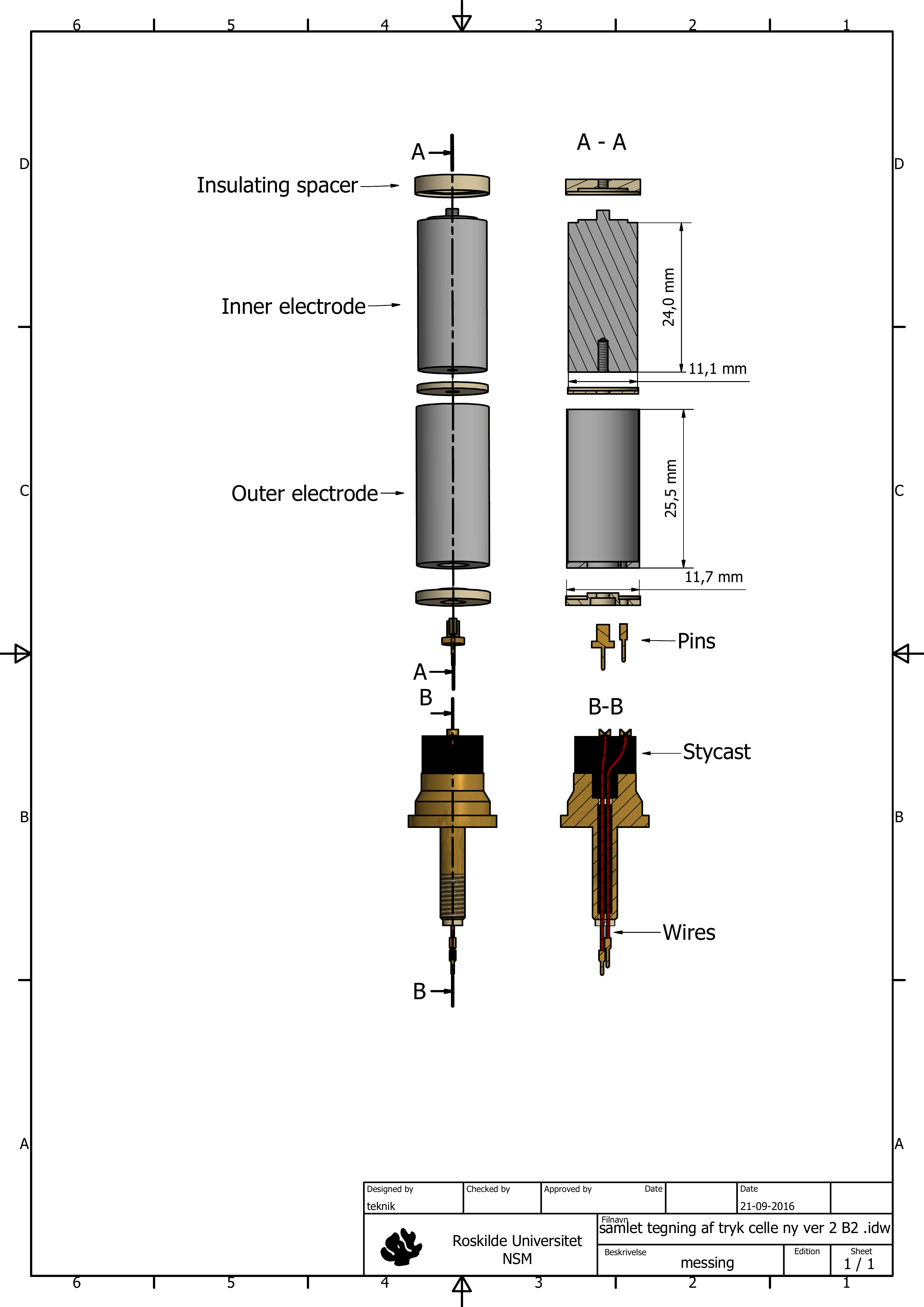}
	\caption{Exploded technical drawing of the dielectric capacitor and sealed plug of the combined neutron scattering and dielectric spectroscopy cell. A cross-sectional description is shown on the right side where the inner and outer electrodes, insulating spacers and pins of the capacitor are described. Cross-sectional view of the sealed plug is also shown.}
	\label{fig:Fig2}
\end{figure}

A detailed drawing of the cylindrical capacitor and sealed plug is presented in Fig.~\ref{fig:Fig2}. The design of the capacitor constitutes a technical challenge since an optimum performance at elevated pressure requires a minimum deformation of the device, avoiding mainly electrode collapse, given the fact that a physical contact between the electrodes would ruin the measurements. The outer electrode is has an inner diameter of 11.4~mm (external diameter 11.7~mm) while the inner electrode has a diameter of 11.1 mm, leading to an internal gap for the sample of 0.15~mm. The inner electrode sits on the bottom insulating spacer which through a cup-like shape allows keeping the outer electrode at a fixed position ensuring the right sample gap. The top and bottom spacers are made of polyether ether ketone (PEEK). Holes in the spacers allow the connection of the electrodes to copper wires and these spacers also serve for insulating the capacitor from the HP container. The height of the capacitor is 24~mm. Simple calculation by applying Gauss' law gives an empty capacitance of 50~pF, in close agreement with the value we found experimentally, approximately 55 $\pm$ 1~pF. This value may change slightly depending on the precise geometry of the capacitor and the exact temperature, but can easily be determined with high precision before each experiment.
 
A key technical issue is how to feed the wires from outside of the sample cell and into the pressurized zone, connecting to the capacitor, while preventing any leak at high pressure and keeping the capacitor electrically isolated from the rest of the cell. For this kind of mechanical sealing and electrical shielding, we developed a plug made of brass with a single cylindrical channel that guides the wires from outside the sample cell and into the pressurized zone. The remaining gap is filled with a dense paste that is then cured to transform into a hard and electrically resistant material. We have used the epoxy encapsulant \textit{STYCAST}\textsuperscript{\textregistered} 2850 FT. A representation of the sealed plug, including the feeding of the wires and the way they are connected to the inner and outer electrodes is shown in Fig.~\ref{fig:Fig2}. Paluch and colleagues proposed a similar solution for dielectric measurements at ultrahigh pressures, but using dental filling as sealing material.\cite{Mierzwa}   

\section{Simultaneous dielectric and neutron spectroscopy measurements at high pressure}

We show some examples of measurements carried out on the BS spectrometer IN16B and on the TOF instrument IN6 at the ILL \cite{Frick,IN16b,IN6}, operating at a neutron wavelength $\lambda$ of 6.27 and 5.18~\AA\ respectively. The high-pressure DS-NS cell was connected to a neutron sample stick specially optimized for this high pressure and dielectric measurements. The stick was designed and developed by the Service for Advanced Neutron Environment (SANE) at the ILL.\cite{Eddy} The sample stick plays three main roles; first, it allows locating the sample in the path of the neutron beam within the calorimeter of a cryostat; second, it drives the pressure-transmitting medium from the compressor to the sample through a capillary, avoiding freezing of the pressure-liquid at the cold point of the cryostat; and third, it brings two coaxial cables that are connected through the plug to the capacitor.
We utilized as pressure controller the electrical pressure multiplicator \textit{Louise} which transmits hydrostatic pressure to the sample with non-hydrogenous liquid Fluorinert\textsuperscript{TM}.\cite{Sidorov} Two sensors measure the pressure, in such a way that the controller may decide to increase or decrease the pressure in order to reach and maintain the desired set-point.\cite{Peters} The pressure controller \textit{Louise} has also been developed at the ILL and can be operated with the instrument control software NOMAD.\cite{Nomad}

The dielectric electronics used for measuring the capacitance of the sample consists of a commercial LCR meter (Agilent E4980A) covering the frequency range $10^{2}-10^{6}$~Hz, while a multimeter in combination with a frequency generator covers the low frequency window $10^{-3}-10^{2}$~Hz.\cite{Igarashi} This particular setup and the MATLAB-based control were developed at Roskilde University and uses an automatic switch to couple the setup at 100~Hz. In order to synchronize neutron and dielectric measurements, an interface of the dielectric software was integrated in NOMAD at the ILL.

\subsection{Experimental details}

In the following lines we explain how the cell and capacitor are filled, as well as the order of assembly. First, the separator or floating barrier is inserted into the high-pressure cell at room temperature and the bore of the HP vessel is then filled with enough sample liquid to ensure a complete fill of the cell after the introduction of the capacitor. For measuring the dielectric signal, the capacitor is filled using capillary force; the gap separating the electrodes of the capacitor is simply filled by placing the assembled capacitor in the sample liquid and letting it get sucked inside by capillary effects. The insulating spacer of the capacitor possesses four gaps for making the filling of the capacitor possible. If the liquid is too viscous, this filling can be accelerated by moderate heating. This creates the electrode-electrode sample compartment. Once the capacitor is filled, it is inserted into the HP cell. The filled gap between the inner wall of the HP vessel and the outer electrode forms the cell-electrode sample compartment. Then, as illustrated in Fig.~\ref{fig:Fig1}, the plug with feedthroughs incorporating rubber and copper seals is connected to the capacitor. The cell is finally locked with the nut and potentially trapped air is let out as it is closed. 
It is important to note that neutrons are in principle scattered from sample located in two environments. A schematic representation of the sample geometry is shown in Fig.~\ref{fig:Fig2}.

\subsection{Examples of simultaneous data collection in dielectric and neutron spectroscopy techniques}
\begin{figure}
	\begin{center}
		\includegraphics[trim = 0mm 0mm 0mm 5mm,clip,width=1.0\linewidth]{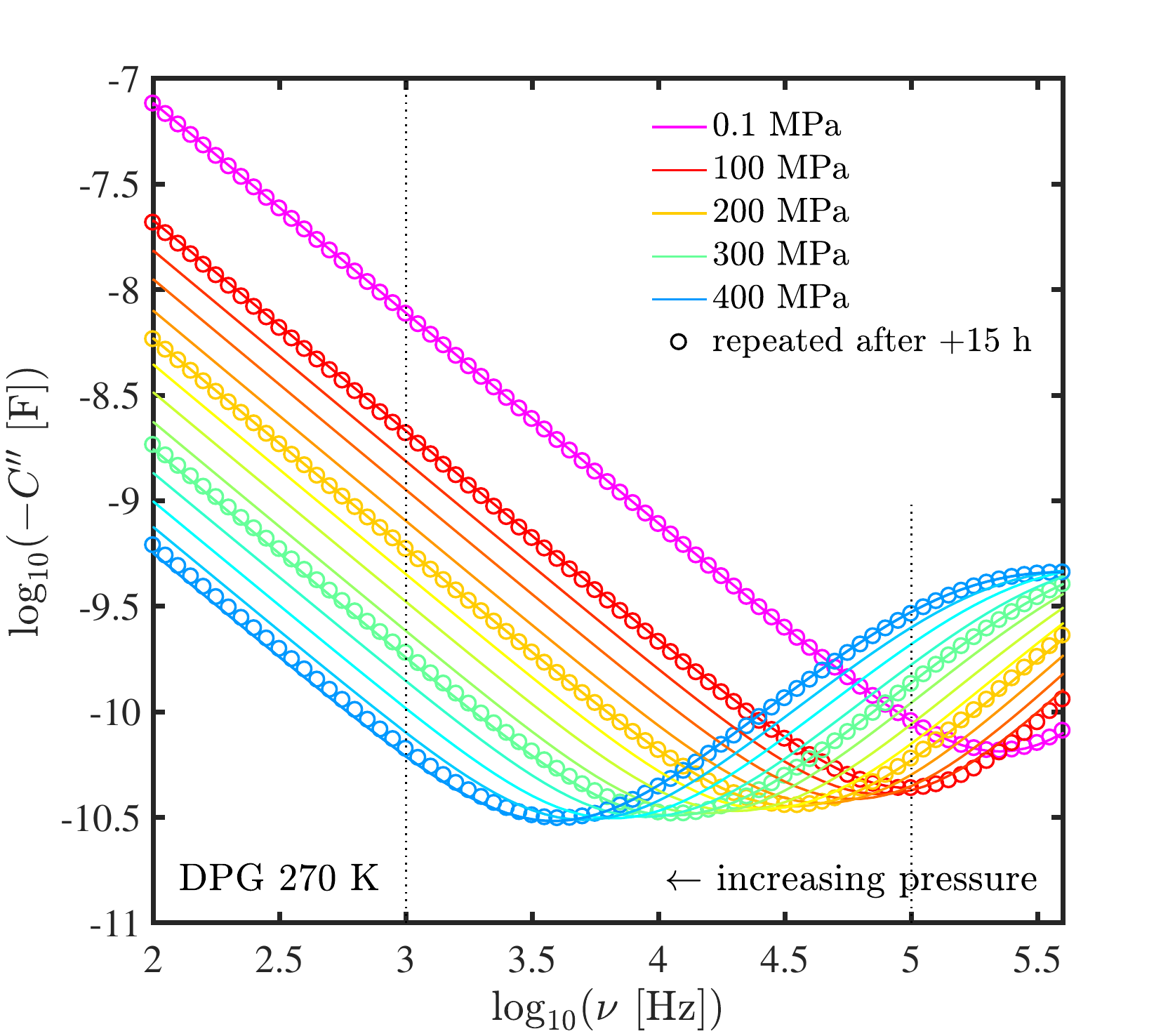}
		\vspace{-0.0cm} 
		\includegraphics[trim = 0mm 0mm 0mm 0mm,clip,width=1\linewidth]{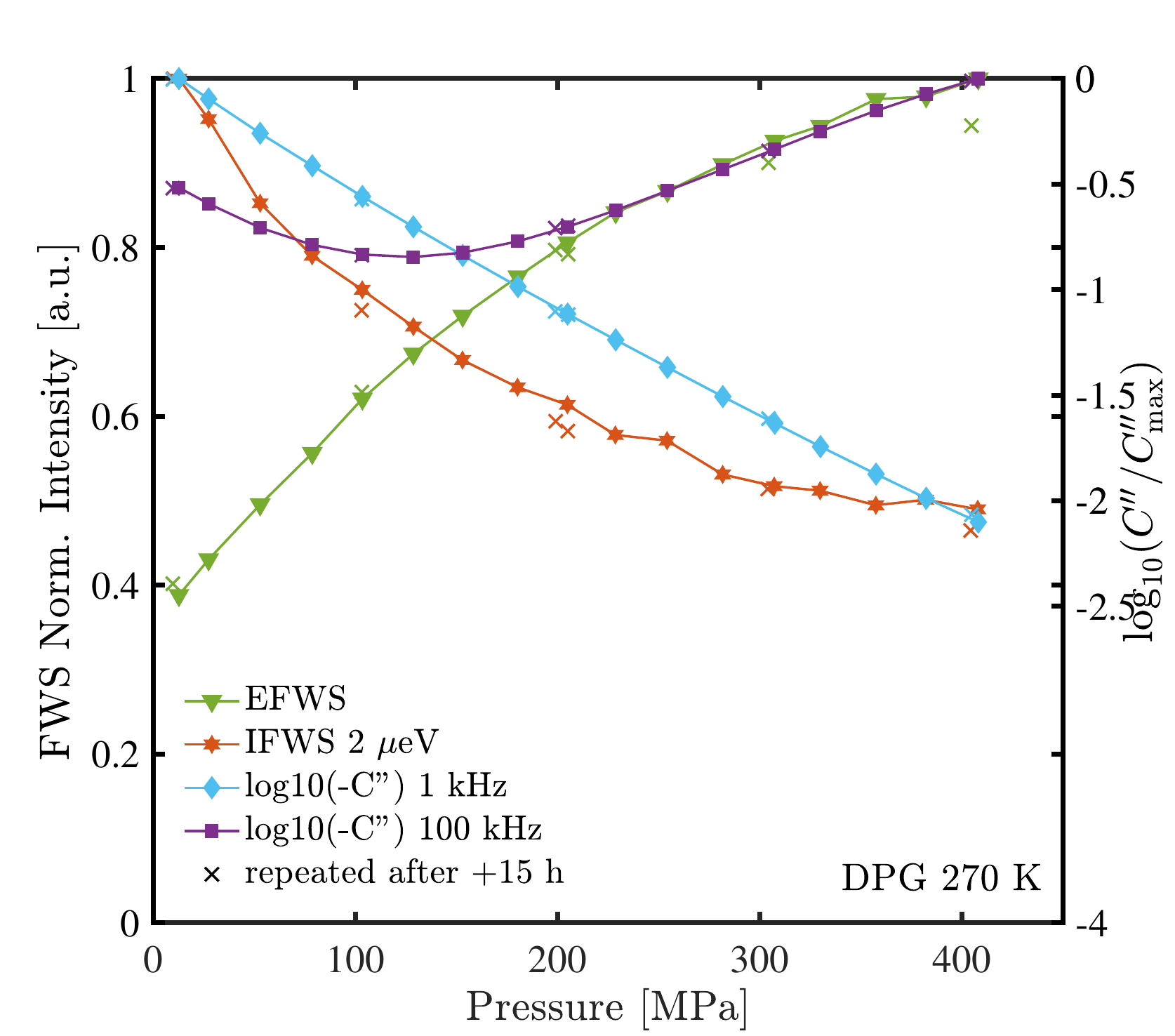}
		\vspace{0.0cm}	
		\caption{Simultaneous dielectric spectroscopy measurements (top), and elastic and inelastic fixed window scans with neutrons (bottom), on DPG as a function of pressure at 270~K. Neutron intensities were acquired on the backscattering instrument IN16B (ILL) and were summed over $\mathbf{Q}$. The bottom panel also shows the pressure dependence of the sample capacitance (imaginary part) at two selected frequencies, 1 and 100~kHz (vertical dotted lines in top panel). Empty symbols (top) and crosses (bottom) correspond to a second pressure scan carried out 15~h later.}
		\label{fig:Fig3}
     \end{center}   
\end{figure}

We present simultaneous pressure and temperature measurements on the molecular glass former dipropylene glycol (DPG). This hydrogen-bonded liquid has a glass transition temperature ($T_{g}$) of 195~K and a pressure coefficient of $T_{g}$ of $dT_{g}/dP=0.08$~K MPa$^{-1}$.\cite{Casalini,Grzybowska} As an example, in Fig.~\ref{fig:Fig3}, we report a pressure scan obtained on IN16B at the ILL, where we recorded simultaneous dielectric spectra and neutron fixed window scans \cite{Frick} in a pressure range between 0.1 and 400~MPa at 270~K. The top panel of Fig.~\ref{fig:Fig3} displays a double logarithmic representation of the imaginary part of the capacitance of DPG as a function of frequency. On the low frequency side, we observe a power law contribution due to free charge conductivity which shifts towards lower frequencies as pressure goes up. On the high frequency side, the low frequency tail of the $\alpha$ relaxation also moves towards lower frequencies with pressure and finally the $\alpha$-peak enters the experimental frequency window. In the bottom panel of Fig.~\ref{fig:Fig3}, we have selected two frequencies to illustrate how we can monitor the change with pressure in conductivity at 1~kHz and in the relaxation of the liquid at 100~kHz (right axis). In the bottom panel of Fig.~\ref{fig:Fig3} we also show the intensity evolution of the elastic and inelastic fixed window scans (EFWS and IFWS, respectively) on increased pressure. EFWS (energy transfer, $\Delta E=0$) have been widely used in different fields, mainly to estimate the mean square displacement through the $\mathbf{Q}$ dependence of the elastic intensity. In contrast, if the incident and scattered neutron wave-vectors are different, $\mathbf{k_{i}} \neq \mathbf{k_{f}}$, but constant during the acquisition process, the technique is referred to as IFWS ($\Delta E\neq0$) and can be alternated with EFWS during pressure changes.\cite{Frick} Here are shown IFWS data for an energy offset of 2~$\mu$eV, together with the EFWS (left axis).
The reduced mobility with pressure produces a progressive increase of the elastic intensity. The inelastic intensity decreases accordingly as the quasielastic signal narrows with the slowing down of the $\alpha$ relaxation in the measured $T,P$-range for DPG and thus its spectral weight decreases at 2~$\mu$eV energy transfer.

Up and down pressure cycles were successfully carried out exhibiting reproducible values of the scattered neutron intensity and sample capacitance, corroborating that the floating barrier generally applies pressure to the sample and is able to retract again to the initial position when pressure is released. As an example, we compare the pressure dependence of dielectric and neutron data between two scans performed 15 h apart, illustrated with $\times$ in Fig.~\ref{fig:Fig3}. One observes that the results are quite similar in both pressure scans.

\subsection{Practical advantages}

\begin{figure}
	\begin{center}
		\includegraphics[trim = 0mm 0mm 5mm 3mm,clip,width=1\linewidth]{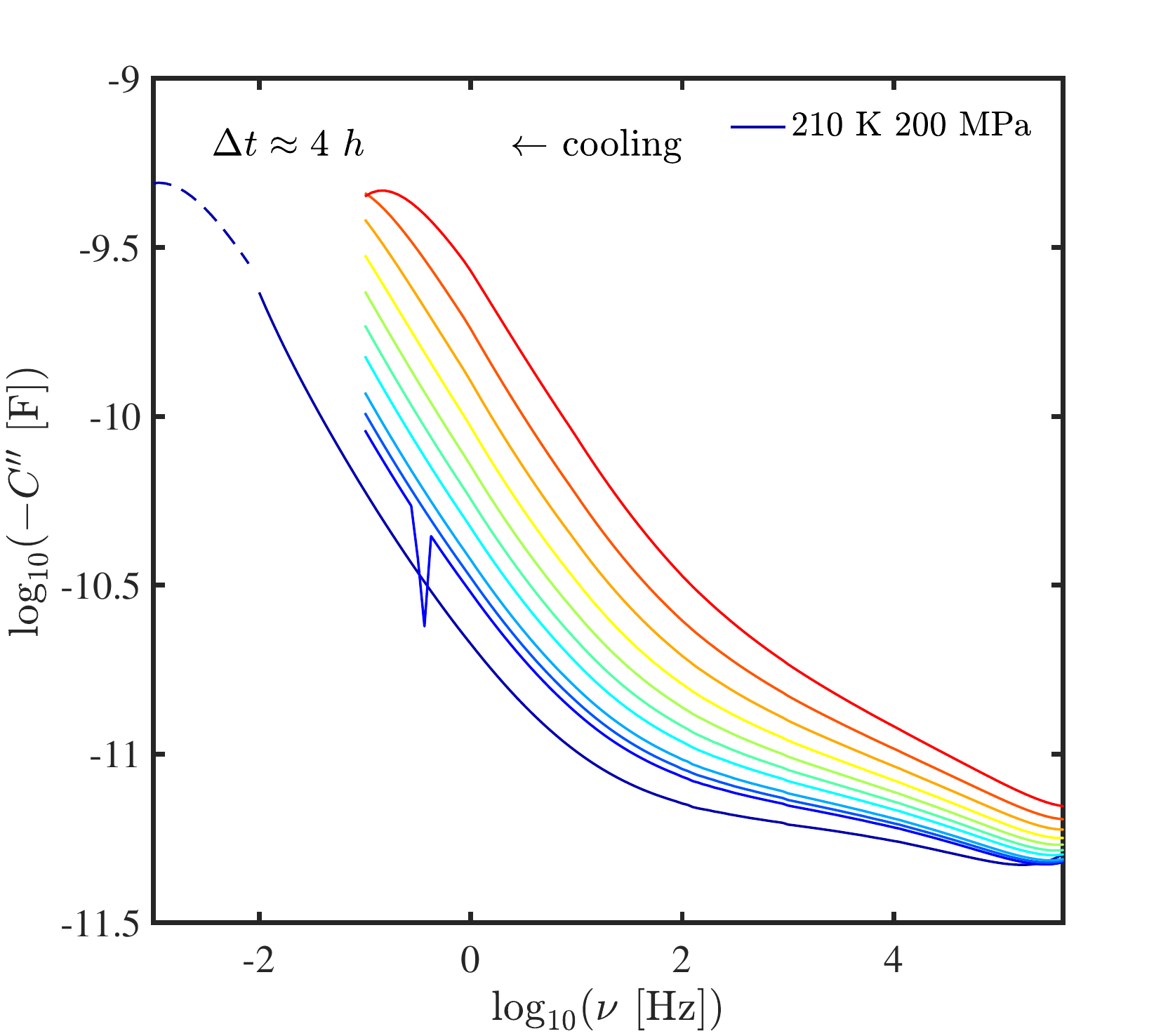}
		\vspace{-0.0cm} 
		\includegraphics[trim = 0mm 0mm 5mm 0mm,clip,width=1\linewidth]{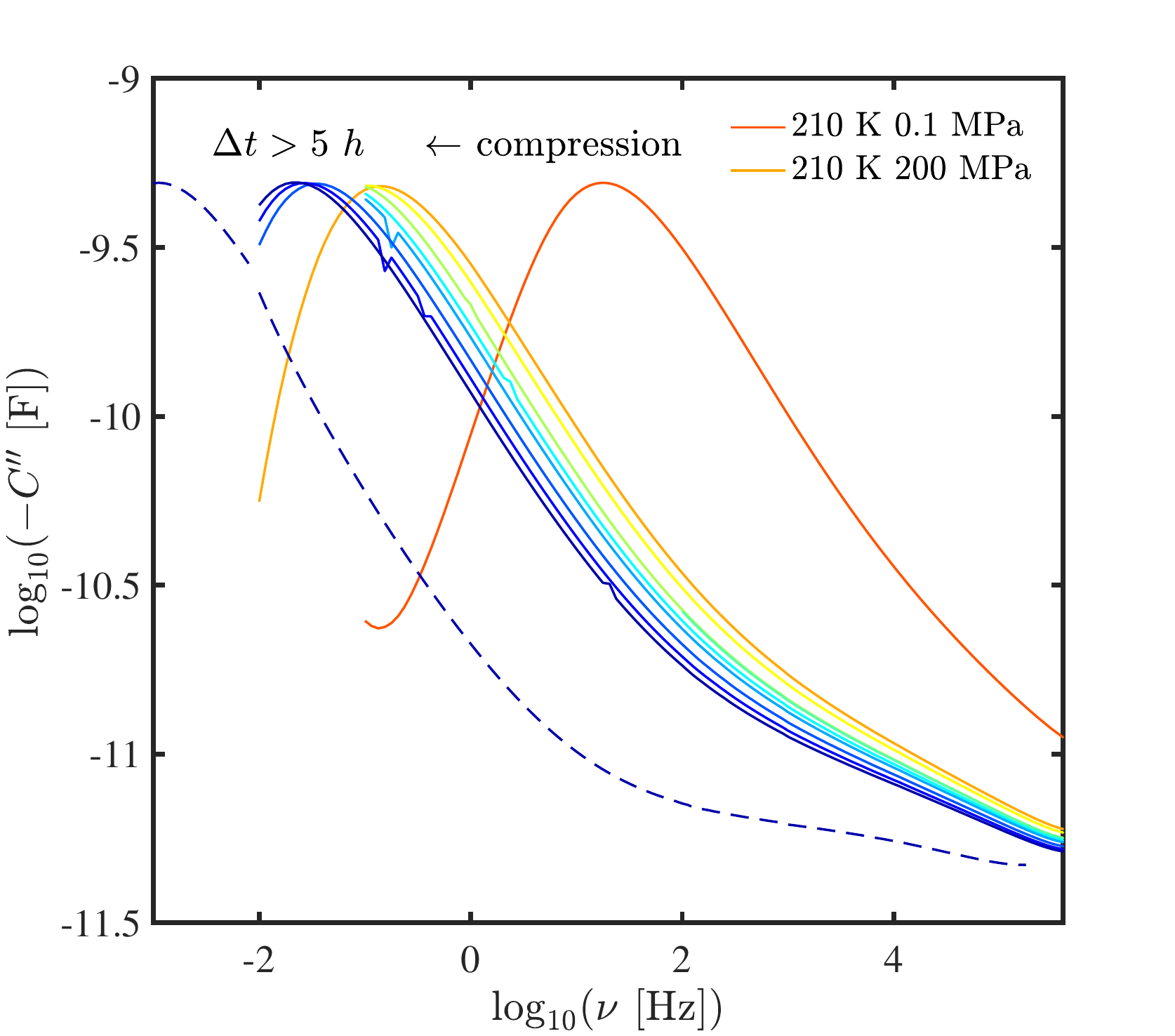}
		\vspace{0.0cm}	
		\caption{Double logarithmic plots of the imaginary part of the capacitance as a function of frequency for DPG when approaching the glass transition either on cooling at constant pressure (top), or compressing at constant temperature (bottom). Both graphs illustrate the sample relaxation towards equilibrium. Dashed line in the top panel at 210~K and 200~MPa show the maximum of the loss peak located at a frequency corresponding to $T_g$ and is obtained through time-temperature superposition (TTS). In the bottom panel, the dashed line corresponds to the expected dielectric spectrum for the equilibrated liquid at the glass transition state point, however, this state point is never reached.}
		\label{fig:Fig5}
	\end{center}   
\end{figure}

The $\alpha$ relaxation is one of the dynamic features that can be probed with dielectric spectroscopy. Due to its extreme sensitivity to pressure and/or temperature changes, the dielectrics acts like an internal probe to define exactly the thermodynamic state of the system. We take advantage of this powerful property to gain information on, for example, whether or not the sample is equilibrated when it approaches the glassy state either by cooling or compression. Figure~\ref{fig:Fig5} (top) shows the dielectric $\alpha$ relaxation of DPG when entering the glassy state at fixed $P=200$~MPa upon cooling down to 210~K. The dielectric spectra from $10^{-1}$~Hz are roughly 10 minutes apart. By reaching the 210~K and 200~MPa state point on cooling at constant pressure (top panel), one observes that the system is mostly thermally equilibrated with a continuous shift of the $\alpha$ relaxation towards lower frequencies. Finally, after approximately 4 hours, dependent on the cooling rate of the cryostat and the heat conductivity of the sample environment, the peak of the $\alpha$ relaxation is located at the characteristic frequency of the glass transition. The maximum (dashed line) is estimated by time-temperature superposition (TTS). If we instead try to reach the glass transition on compression at constant $T=210$~K (bottom panel), there is an immediate response to the pressure change but then the $\alpha$ relaxation is practically \textit{stuck} at a frequency of the maximum loss around 0.05~Hz, indicating that the sample or the pressure liquid becomes too viscous for efficiently transmitting the pressure from the floating barrier. On compression (bottom panel), the three last measurements in blue from $10^{-2}$~Hz take 1~h each and show practically no movement of the maximum loss peak. This aspect is critical for viscous liquids, in which the structural relaxation is extremely dependent on temperature and pressure, that is, the $\alpha$ relaxation time often changes with an order of magnitude as a response to small changes in temperature and/or pressure.\cite{Angell,Berthier} 

\begin{figure}
	
	\includegraphics[trim = 0mm 0mm 10mm 5mm, clip, width=1.0\linewidth]{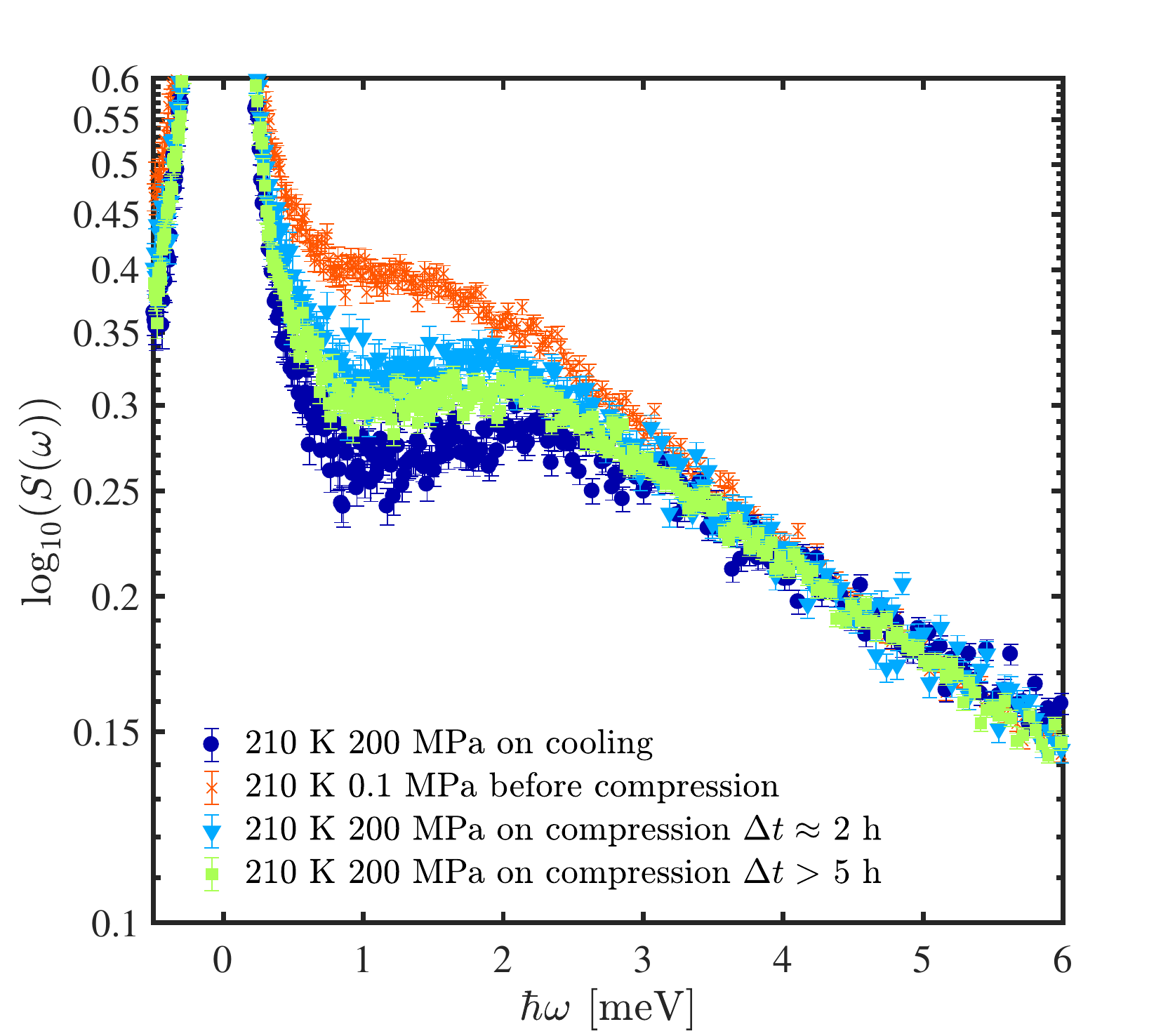}
	\caption{Spectra summed over all \textit{Q} from IN6 of DPG at the equilibrated state point 210~K 200~MPa reached on cooling (state point corresponding to the last dielectric frame included in the top panel of Fig. \ref{fig:Fig5}), and the compression from 210~K 0.1~MPa to 200 MPa at different equilibration times (corresponding to dielectric data included in the bottom panel of Fig. \ref{fig:Fig5}). The vertical axis is shown up to a 0.4 $\%$ of the maximum intensity of the central elastic peak.}
	\label{fig:Fig6}
\end{figure}

The fact that pressure is sometimes not transmitted properly onto the sample volume when entering the glass on compression is illustrated by the simultaneous DS and NS measurements. Neutron TOF measurements from IN6 are presented in Fig.~\ref{fig:Fig6}, and where carried out simultaneously with the dielectric spectra in Fig.~\ref{fig:Fig5}. In Fig.~\ref{fig:Fig6} we compare the dynamic structure factor summed over $\mathbf{Q}$, $S(\omega)$, of DPG at 210~K and 200~MPa coming into the glass on compression or on cooling. The spectra have been grouped for constant values of $\mathbf{Q}$ in the range 1.2-2~\AA$^{-1}$. The spectrum on compression after a waiting time of more than 5 hours does not match the curve shown by the equilibrated glass formed on cooling, which is in agreement with the simultaneous dielectric data shown in Fig.~\ref{fig:Fig5}. Unlike the long acquisition periods required to obtain well-resolved TOF spectra (up to a couple of hours, depending on the neutron flux), full dielectric curves from 100~mHz to 1~MHz can be measured in 10 minutes approximately. DS offers quick and precise information about whether or not the sample is in thermodynamic equilibrium and represents a significant advantage compared to standard neutron techniques. These results indicate that without the knowledge from simultaneous DS one might erroneously interpret the changes in $S(\omega)$ being due to a path dependence in the $(P,T)$ phase diagram.

The dielectric signal from the combined cell also provides fast information if a sample is prone to crystallization, and also if mixing of pressure liquid and sample occurs. The HP combined cell provides information immediately if there is a leak or pressure-transmitting liquid enters the sample environment. In some rare cases the plug started to leak, in particular when the plug was subjected to several cycles up to 400~MPa. Leakage in the plug is presumably induced by formation of cracks on the sealing epoxy resin used for feedthroughs. In any case, the failure of the plug is certainly unpredictable since several factors could contribute to the loss of tightness. 

Another practical benefit of coupling neutron with dielectric spectroscopy, in one of the measurements on DPG at 400~MPa, we detected a strange and very strong conductivity signal entering the dielectric spectrum at low frequencies. We concluded that Fluorinert\textsuperscript{TM}, the pressure-transmitting liquid, was entering in the sample compartment due to loss of floating barrier tightness, an example of a scenario which is difficult to detect with neutrons alone.

\section{Conclusions}

We have designed, constructed and tested an experimental setup to perform simultaneous neutron and dielectric spectroscopy experiments at high pressure. 
The performance of the cell has been successfully proved in a broad temperature range and up to a maximum pressure of 500~MPa. It opens new possibilities to address studies on dynamic processes in an extremely broad range of timescales. Moreover, measuring the dielectric signal of the sample is an excellent internal probe to get insight into the state of thermodynamic equilibrium after temperature and pressure changes.

\begin{acknowledgments}
This work has been funded by the Danish Council for Independent Research (Sapere Aude: Starting Grant). The authors gratefully acknowledge use of the facilities at the Institut Laue-Langevin (Grenoble, France) through the Long Term Proposal LTP-6-7. The authors would like to thank the support from Jean Paul Gonzales and  Claude Payre (Service for Advanced Neutron Environment at the ILL) and S\'ebastian Vial (Diffraction Group at the ILL). We thank Preben Olsen, Ebbe H. Larsen  and Torben Rasmussen (workshop at IMFUFA, Department of Science and Environment, Roskilde University) for their support on the design and manufacture of the high-pressure cell. Christiane Alba-Simionesco is also acknowledged for fruitful discussions.  
\end{acknowledgments}


\begin{thebibliography}{30}%
	\makeatletter
	\providecommand \@ifxundefined [1]{%
		\@ifx{#1\undefined}
	}%
	\providecommand \@ifnum [1]{%
		\ifnum #1\expandafter \@firstoftwo
		\else \expandafter \@secondoftwo
		\fi
	}%
	\providecommand \@ifx [1]{%
		\ifx #1\expandafter \@firstoftwo
		\else \expandafter \@secondoftwo
		\fi
	}%
	\providecommand \natexlab [1]{#1}%
	\providecommand \enquote  [1]{``#1''}%
	\providecommand \bibnamefont  [1]{#1}%
	\providecommand \bibfnamefont [1]{#1}%
	\providecommand \citenamefont [1]{#1}%
	\providecommand \href@noop [0]{\@secondoftwo}%
	\providecommand \href [0]{\begingroup \@sanitize@url \@href}%
	\providecommand \@href[1]{\@@startlink{#1}\@@href}%
	\providecommand \@@href[1]{\endgroup#1\@@endlink}%
	\providecommand \@sanitize@url [0]{\catcode `\\12\catcode `\$12\catcode
		`\&12\catcode `\#12\catcode `\^12\catcode `\_12\catcode `\%12\relax}%
	\providecommand \@@startlink[1]{}%
	\providecommand \@@endlink[0]{}%
	\providecommand \url  [0]{\begingroup\@sanitize@url \@url }%
	\providecommand \@url [1]{\endgroup\@href {#1}{\urlprefix }}%
	\providecommand \urlprefix  [0]{URL }%
	\providecommand \Eprint [0]{\href }%
	\providecommand \doibase [0]{http://dx.doi.org/}%
	\providecommand \selectlanguage [0]{\@gobble}%
	\providecommand \bibinfo  [0]{\@secondoftwo}%
	\providecommand \bibfield  [0]{\@secondoftwo}%
	\providecommand \translation [1]{[#1]}%
	\providecommand \BibitemOpen [0]{}%
	\providecommand \bibitemStop [0]{}%
	\providecommand \bibitemNoStop [0]{.\EOS\space}%
	\providecommand \EOS [0]{\spacefactor3000\relax}%
	\providecommand \BibitemShut  [1]{\csname bibitem#1\endcsname}%
	\let\auto@bib@innerbib\@empty
	\bibitem [{\citenamefont {Ediger}, \citenamefont {Angell},\ and\ \citenamefont
		{Nagel}(1996)}]{Ediger}%
	\BibitemOpen
	\bibfield  {author} {\bibinfo {author} {\bibfnamefont {M.}~\bibnamefont
			{Ediger}}, \bibinfo {author} {\bibfnamefont {C.}~\bibnamefont {Angell}}, \
		and\ \bibinfo {author} {\bibfnamefont {S.}~\bibnamefont {Nagel}},\
	}\href@noop {} {\bibfield  {journal} {\bibinfo  {journal} {Journal of
				Physical Chemistry}\ }\textbf {\bibinfo {volume} {100}},\ \bibinfo {pages}
		{13200} (\bibinfo {year} {1996})}\BibitemShut {NoStop}%
	\bibitem [{\citenamefont {Berthier}\ and\ \citenamefont
		{Ediger}(2016)}]{Berthier2}%
	\BibitemOpen
	\bibfield  {author} {\bibinfo {author} {\bibfnamefont {L.}~\bibnamefont
			{Berthier}}\ and\ \bibinfo {author} {\bibfnamefont {M.~D.}\ \bibnamefont
			{Ediger}},\ }\href@noop {} {\bibfield  {journal} {\bibinfo  {journal}
			{Physics Today}\ }\textbf {\bibinfo {volume} {69}},\ \bibinfo {pages} {40}
		(\bibinfo {year} {2016})}\BibitemShut {NoStop}%
	\bibitem [{\citenamefont {Roland}\ \emph {et~al.}(2005)\citenamefont {Roland},
		\citenamefont {Hensel-Bielowka}, \citenamefont {Paluch},\ and\ \citenamefont
		{Casalini}}]{Roland}%
	\BibitemOpen
	\bibfield  {author} {\bibinfo {author} {\bibfnamefont {C.~M.}\ \bibnamefont
			{Roland}}, \bibinfo {author} {\bibfnamefont {S.}~\bibnamefont
			{Hensel-Bielowka}}, \bibinfo {author} {\bibfnamefont {M.}~\bibnamefont
			{Paluch}}, \ and\ \bibinfo {author} {\bibfnamefont {R.}~\bibnamefont
			{Casalini}},\ }\href@noop {} {\bibfield  {journal} {\bibinfo  {journal}
			{Reports on Progress in Physics}\ }\textbf {\bibinfo {volume} {68}},\
		\bibinfo {pages} {1405} (\bibinfo {year} {2005})}\BibitemShut {NoStop}%
	\bibitem [{\citenamefont {Pazmi\~no Betancourt}\ \emph
		{et~al.}(2015)\citenamefont {Pazmi\~no Betancourt}, \citenamefont {Hanakata},
		\citenamefont {Starr},\ and\ \citenamefont {Douglas}}]{Betancourt}%
	\BibitemOpen
	\bibfield  {author} {\bibinfo {author} {\bibfnamefont {B.~A.}\ \bibnamefont
			{Pazmi\~no Betancourt}}, \bibinfo {author} {\bibfnamefont {P.~Z.}\
			\bibnamefont {Hanakata}}, \bibinfo {author} {\bibfnamefont {F.~W.}\
			\bibnamefont {Starr}}, \ and\ \bibinfo {author} {\bibfnamefont {J.~F.}\
			\bibnamefont {Douglas}},\ }\href {\doibase 10.1073/pnas.1418654112}
	{\bibfield  {journal} {\bibinfo  {journal} {Proceedings of the National
				Academy of Sciences}\ }\textbf {\bibinfo {volume} {112}},\ \bibinfo {pages}
		{2966} (\bibinfo {year} {2015})}\BibitemShut {NoStop}%
	\bibitem [{\citenamefont {Hansen}\ \emph {et~al.}(2017)\citenamefont {Hansen},
		\citenamefont {Frick}, \citenamefont {Hecksher}, \citenamefont {Dyre},\ and\
		\citenamefont {Niss}}]{Henriette}%
	\BibitemOpen
	\bibfield  {author} {\bibinfo {author} {\bibfnamefont {H.~W.}\ \bibnamefont
			{Hansen}}, \bibinfo {author} {\bibfnamefont {B.}~\bibnamefont {Frick}},
		\bibinfo {author} {\bibfnamefont {T.}~\bibnamefont {Hecksher}}, \bibinfo
		{author} {\bibfnamefont {J.~C.}\ \bibnamefont {Dyre}}, \ and\ \bibinfo
		{author} {\bibfnamefont {K.}~\bibnamefont {Niss}},\ }\href {\doibase
		10.1103/PhysRevB.95.104202} {\bibfield  {journal} {\bibinfo  {journal}
			{Physical Review B}\ }\textbf {\bibinfo {volume} {95}},\ \bibinfo {pages}
		{104202} (\bibinfo {year} {2017})}\BibitemShut {NoStop}%
	\bibitem [{\citenamefont {Sch\"{o}nhals}\ and\ \citenamefont
		{Kremer}(2003)}]{Kremer}%
	\BibitemOpen
	\bibfield  {author} {\bibinfo {author} {\bibfnamefont {A.}~\bibnamefont
			{Sch\"{o}nhals}}\ and\ \bibinfo {author} {\bibfnamefont {F.}~\bibnamefont
			{Kremer}},\ }\href@noop {} {\bibfield  {journal} {\bibinfo  {journal}
			{Broadband Dielectric Spectroscopy, Springer-Verlag Berlin Heidelberg}\ }
		(\bibinfo {year} {2003})}\BibitemShut {NoStop}%
	\bibitem [{\citenamefont {B\'ee}(1988)}]{Bee}%
	\BibitemOpen
	\bibfield  {author} {\bibinfo {author} {\bibfnamefont {M.}~\bibnamefont
			{B\'ee}},\ }\href@noop {} {\emph {\bibinfo {title} {Quasielastic neutron
				scattering}}}\ (\bibinfo  {publisher} {United Kingdom: Adam Hilger},\
	\bibinfo {year} {1988})\BibitemShut {NoStop}%
	\bibitem [{\citenamefont {Lovesey}(1984)}]{Lovesey}%
	\BibitemOpen
	\bibfield  {author} {\bibinfo {author} {\bibfnamefont {S.}~\bibnamefont
			{Lovesey}},\ }\href {https://books.google.dk/books?id=xfLvAAAAMAAJ} {\emph
		{\bibinfo {title} {Theory of neutron scattering from condensed matter}}},\
	\bibinfo {series} {International series of monographs on physics}\ No.\
	\bibinfo {number} {v. 1}\ (\bibinfo  {publisher} {Clarendon Press},\ \bibinfo
	{year} {1984})\BibitemShut {NoStop}%
	\bibitem [{\citenamefont {Springer}\ and\ \citenamefont
		{Lechner}(2005)}]{Springer}%
	\BibitemOpen
	\bibfield  {author} {\bibinfo {author} {\bibfnamefont {T.}~\bibnamefont
			{Springer}}\ and\ \bibinfo {author} {\bibfnamefont {R.~E.}\ \bibnamefont
			{Lechner}},\ }\enquote {\bibinfo {title} {Diffusion studies of solids by
			quasielastic neutron scattering},}\ in\ \href@noop {} {\emph {\bibinfo
			{booktitle} {Diffusion in Condensed Matter: Methods, Materials, Models}}},\
	\bibinfo {editor} {edited by\ \bibinfo {editor} {\bibfnamefont
			{P.}~\bibnamefont {Heitjans}}\ and\ \bibinfo {editor} {\bibfnamefont
			{J.}~\bibnamefont {K{\"a}rger}}}\ (\bibinfo  {publisher} {Springer Berlin
		Heidelberg},\ \bibinfo {address} {Berlin, Heidelberg},\ \bibinfo {year}
	{2005})\ pp.\ \bibinfo {pages} {93--164}\BibitemShut {NoStop}%
	\bibitem [{\citenamefont {Debenedetti}\ and\ \citenamefont
		{Stillinger}(2001)}]{Debenedetti}%
	\BibitemOpen
	\bibfield  {author} {\bibinfo {author} {\bibfnamefont {P.}~\bibnamefont
			{Debenedetti}}\ and\ \bibinfo {author} {\bibfnamefont {F.}~\bibnamefont
			{Stillinger}},\ }\href@noop {} {\bibfield  {journal} {\bibinfo  {journal}
			{Nature}\ }\textbf {\bibinfo {volume} {410}},\ \bibinfo {pages} {259}
		(\bibinfo {year} {2001})}\BibitemShut {NoStop}%
	\bibitem [{\citenamefont {Colmenero}(2015)}]{Colmenero}%
	\BibitemOpen
	\bibfield  {author} {\bibinfo {author} {\bibfnamefont {J.}~\bibnamefont
			{Colmenero}},\ }\href@noop {} {\bibfield  {journal} {\bibinfo  {journal}
			{Journal of Physics: Condensed Matter}\ }\textbf {\bibinfo {volume} {27}},\
		\bibinfo {pages} {103101} (\bibinfo {year} {2015})}\BibitemShut {NoStop}%
	\bibitem [{\citenamefont {Klotz}(2012)}]{Klotz}%
	\BibitemOpen
	\bibfield  {author} {\bibinfo {author} {\bibfnamefont {S.}~\bibnamefont
			{Klotz}},\ }\href {https://books.google.dk/books?id=sqDFAwJ3C\_oC} {\emph
		{\bibinfo {title} {Techniques in High Pressure Neutron Scattering}}}\
	(\bibinfo  {publisher} {Taylor \& Francis},\ \bibinfo {year}
	{2012})\BibitemShut {NoStop}%
	\bibitem [{IN6()}]{IN6}%
	\BibitemOpen
	\href@noop {} {\bibinfo  {journal}
		{http://www.ill.eu/instruments-support/instruments-groups/ instruments/in6/}\
	}\BibitemShut {NoStop}%
	\bibitem [{\citenamefont {Frick}, \citenamefont {Combet},\ and\ \citenamefont
		{van Eijck}(2012)}]{Frick}%
	\BibitemOpen
	\bibfield  {journal} {  }\bibfield  {author} {\bibinfo {author} {\bibfnamefont
			{B.}~\bibnamefont {Frick}}, \bibinfo {author} {\bibfnamefont
			{J.}~\bibnamefont {Combet}}, \ and\ \bibinfo {author} {\bibfnamefont
			{L.}~\bibnamefont {van Eijck}},\ }\href@noop {} {\bibfield  {journal}
		{\bibinfo  {journal} {Nuclear Instruments and Methods in Physics Research
				Section A: Accelerators, Spectrometers, Detectors and Associated Equipment}\
		}\textbf {\bibinfo {volume} {669}},\ \bibinfo {pages} {7} (\bibinfo {year}
		{2012})}\BibitemShut {NoStop}%
	\bibitem [{IN1()}]{IN16b}%
	\BibitemOpen
	\href@noop {} {\bibinfo  {journal}
		{https://www.ill.eu/instruments-support/instruments-groups/instruments/in16b/description/instrument-layout/}\
	}\BibitemShut {NoStop}%
	\bibitem [{\citenamefont {Natali}\ \emph {et~al.}(2008)\citenamefont {Natali},
		\citenamefont {Peters}, \citenamefont {Russo}, \citenamefont {Barbieri},
		\citenamefont {Chiapponi}, \citenamefont {Cupane}, \citenamefont {Deriu},
		\citenamefont {Bari}, \citenamefont {Farhi}, \citenamefont {Gerelli},
		\citenamefont {Mariani}, \citenamefont {Paciaroni}, \citenamefont
		{Rivasseau}, \citenamefont {Schirò},\ and\ \citenamefont
		{Sonvico}}]{Natali}%
	\BibitemOpen
	\bibfield  {journal} {  }\bibfield  {author} {\bibinfo {author} {\bibfnamefont
			{F.}~\bibnamefont {Natali}}, \bibinfo {author} {\bibfnamefont
			{J.}~\bibnamefont {Peters}}, \bibinfo {author} {\bibfnamefont
			{D.}~\bibnamefont {Russo}}, \bibinfo {author} {\bibfnamefont
			{S.}~\bibnamefont {Barbieri}}, \bibinfo {author} {\bibfnamefont
			{C.}~\bibnamefont {Chiapponi}}, \bibinfo {author} {\bibfnamefont
			{A.}~\bibnamefont {Cupane}}, \bibinfo {author} {\bibfnamefont
			{A.}~\bibnamefont {Deriu}}, \bibinfo {author} {\bibfnamefont {M.~T.~D.}\
			\bibnamefont {Bari}}, \bibinfo {author} {\bibfnamefont {E.}~\bibnamefont
			{Farhi}}, \bibinfo {author} {\bibfnamefont {Y.}~\bibnamefont {Gerelli}},
		\bibinfo {author} {\bibfnamefont {P.}~\bibnamefont {Mariani}}, \bibinfo
		{author} {\bibfnamefont {A.}~\bibnamefont {Paciaroni}}, \bibinfo {author}
		{\bibfnamefont {C.}~\bibnamefont {Rivasseau}}, \bibinfo {author}
		{\bibfnamefont {G.}~\bibnamefont {Schirò}}, \ and\ \bibinfo {author}
		{\bibfnamefont {F.}~\bibnamefont {Sonvico}},\ }\href {\doibase
		10.1080/10448630802474083} {\bibfield  {journal} {\bibinfo  {journal}
			{Neutron News}\ }\textbf {\bibinfo {volume} {19}},\ \bibinfo {pages} {14}
		(\bibinfo {year} {2008})}\BibitemShut {NoStop}%
	\bibitem [{\citenamefont {Frick}\ and\ \citenamefont
		{Alba-Simionesco}(2002)}]{Frick2002}%
	\BibitemOpen
	\bibfield  {author} {\bibinfo {author} {\bibfnamefont {B.}~\bibnamefont
			{Frick}}\ and\ \bibinfo {author} {\bibfnamefont {C.}~\bibnamefont
			{Alba-Simionesco}},\ }\href@noop {} {\bibfield  {journal} {\bibinfo
			{journal} {Applied Physics A}\ }\textbf {\bibinfo {volume} {74}},\ \bibinfo
		{pages} {S549} (\bibinfo {year} {2002})}\BibitemShut {NoStop}%
	\bibitem [{\citenamefont {Kuhs}, \citenamefont {Hensel},\ and\ \citenamefont
		{Bartels}(2005)}]{Kuhs}%
	\BibitemOpen
	\bibfield  {author} {\bibinfo {author} {\bibfnamefont {W.~F.}\ \bibnamefont
			{Kuhs}}, \bibinfo {author} {\bibfnamefont {E.}~\bibnamefont {Hensel}}, \ and\
		\bibinfo {author} {\bibfnamefont {H.}~\bibnamefont {Bartels}},\ }\href@noop
	{} {\bibfield  {journal} {\bibinfo  {journal} {Journal of Physics: Condensed
				Matter}\ }\textbf {\bibinfo {volume} {17}},\ \bibinfo {pages} {S3009}
		(\bibinfo {year} {2005})}\BibitemShut {NoStop}%
	\bibitem [{\citenamefont {Wang}\ \emph {et~al.}(2011)\citenamefont {Wang},
		\citenamefont {Sokolov}, \citenamefont {Huxley},\ and\ \citenamefont
		{Kamenev}}]{Wang}%
	\BibitemOpen
	\bibfield  {author} {\bibinfo {author} {\bibfnamefont {W.}~\bibnamefont
			{Wang}}, \bibinfo {author} {\bibfnamefont {D.~A.}\ \bibnamefont {Sokolov}},
		\bibinfo {author} {\bibfnamefont {A.~D.}\ \bibnamefont {Huxley}}, \ and\
		\bibinfo {author} {\bibfnamefont {K.~V.}\ \bibnamefont {Kamenev}},\ }\href
	{\doibase 10.1063/1.3608112} {\bibfield  {journal} {\bibinfo  {journal}
			{Review of Scientific Instruments}\ }\textbf {\bibinfo {volume} {82}},\
		\bibinfo {pages} {073903} (\bibinfo {year} {2011})}\BibitemShut {NoStop}%
	\bibitem [{\citenamefont {Yang}\ \emph {et~al.}(2011)\citenamefont {Yang},
		\citenamefont {Kaplonski}, \citenamefont {Unruh}, \citenamefont {Mamontov},\
		and\ \citenamefont {Meyer}}]{Yang}%
	\BibitemOpen
	\bibfield  {author} {\bibinfo {author} {\bibfnamefont {F.}~\bibnamefont
			{Yang}}, \bibinfo {author} {\bibfnamefont {J.}~\bibnamefont {Kaplonski}},
		\bibinfo {author} {\bibfnamefont {T.}~\bibnamefont {Unruh}}, \bibinfo
		{author} {\bibfnamefont {E.}~\bibnamefont {Mamontov}}, \ and\ \bibinfo
		{author} {\bibfnamefont {A.}~\bibnamefont {Meyer}},\ }\href {\doibase
		10.1063/1.3623796} {\bibfield  {journal} {\bibinfo  {journal} {Review of
				Scientific Instruments}\ }\textbf {\bibinfo {volume} {82}},\ \bibinfo {pages}
		{083903} (\bibinfo {year} {2011})}\BibitemShut {NoStop}%
	\bibitem [{\citenamefont {Peters}\ \emph {et~al.}(2012)\citenamefont {Peters},
		\citenamefont {Trapp}, \citenamefont {Hughes}, \citenamefont {Rowe},
		\citenamefont {Dem\'e}, \citenamefont {Laborier}, \citenamefont {Payre},
		\citenamefont {Gonzales}, \citenamefont {Baudoin}, \citenamefont {Belkhier},\
		and\ \citenamefont {Leli\`evre-Berna}}]{Peters}%
	\BibitemOpen
	\bibfield  {author} {\bibinfo {author} {\bibfnamefont {J.}~\bibnamefont
			{Peters}}, \bibinfo {author} {\bibfnamefont {M.}~\bibnamefont {Trapp}},
		\bibinfo {author} {\bibfnamefont {D.}~\bibnamefont {Hughes}}, \bibinfo
		{author} {\bibfnamefont {S.}~\bibnamefont {Rowe}}, \bibinfo {author}
		{\bibfnamefont {B.}~\bibnamefont {Dem\'e}}, \bibinfo {author} {\bibfnamefont
			{J.-L.}\ \bibnamefont {Laborier}}, \bibinfo {author} {\bibfnamefont
			{C.}~\bibnamefont {Payre}}, \bibinfo {author} {\bibfnamefont {J.-P.}\
			\bibnamefont {Gonzales}}, \bibinfo {author} {\bibfnamefont {S.}~\bibnamefont
			{Baudoin}}, \bibinfo {author} {\bibfnamefont {N.}~\bibnamefont {Belkhier}}, \
		and\ \bibinfo {author} {\bibfnamefont {E.}~\bibnamefont {Leli\`evre-Berna}},\
	}\href@noop {} {\bibfield  {journal} {\bibinfo  {journal} {High Pressure
				Research}\ }\textbf {\bibinfo {volume} {32}},\ \bibinfo {pages} {97}
		(\bibinfo {year} {2012})}\BibitemShut {NoStop}%
	\bibitem [{\citenamefont {Leli\`evre-Berna}\ \emph {et~al.}(2017)\citenamefont
		{Leli\`evre-Berna}, \citenamefont {Dem\'e}, \citenamefont {Gonthier},
		\citenamefont {Gonzales}, \citenamefont {Maurice}, \citenamefont {Memphis},
		\citenamefont {Payre}, \citenamefont {Oger}, \citenamefont {Peters},\ and\
		\citenamefont {Vial}}]{Eddy}%
	\BibitemOpen
	\bibfield  {author} {\bibinfo {author} {\bibfnamefont {E.}~\bibnamefont
			{Leli\`evre-Berna}}, \bibinfo {author} {\bibfnamefont {B.}~\bibnamefont
			{Dem\'e}}, \bibinfo {author} {\bibfnamefont {J.}~\bibnamefont {Gonthier}},
		\bibinfo {author} {\bibfnamefont {J.-P.}\ \bibnamefont {Gonzales}}, \bibinfo
		{author} {\bibfnamefont {J.}~\bibnamefont {Maurice}}, \bibinfo {author}
		{\bibfnamefont {Y.}~\bibnamefont {Memphis}}, \bibinfo {author} {\bibfnamefont
			{C.}~\bibnamefont {Payre}}, \bibinfo {author} {\bibfnamefont
			{P.}~\bibnamefont {Oger}}, \bibinfo {author} {\bibfnamefont {J.}~\bibnamefont
			{Peters}}, \ and\ \bibinfo {author} {\bibfnamefont {S.}~\bibnamefont
			{Vial}},\ }\href@noop {} {\bibfield  {journal} {\bibinfo  {journal} {Journal
				of Neutron Research}\ }\textbf {\bibinfo {volume} {19}},\ \bibinfo {pages}
		{77–84} (\bibinfo {year} {2017})}\BibitemShut {NoStop}%
	\bibitem [{\citenamefont {Mierzwa}\ \emph {et~al.}(2010)\citenamefont
		{Mierzwa}, \citenamefont {Pawlus}, \citenamefont {Paluch}, \citenamefont
		{Ziolo},\ and\ \citenamefont {Szulc}}]{Mierzwa}%
	\BibitemOpen
	\bibfield  {author} {\bibinfo {author} {\bibfnamefont {M.}~\bibnamefont
			{Mierzwa}}, \bibinfo {author} {\bibfnamefont {S.}~\bibnamefont {Pawlus}},
		\bibinfo {author} {\bibfnamefont {M.}~\bibnamefont {Paluch}}, \bibinfo
		{author} {\bibfnamefont {J.}~\bibnamefont {Ziolo}}, \ and\ \bibinfo {author}
		{\bibfnamefont {A.}~\bibnamefont {Szulc}},\ }\href {\doibase
		10.1063/1.3436465} {\bibfield  {journal} {\bibinfo  {journal} {Review of
				Scientific Instruments}\ }\textbf {\bibinfo {volume} {81}},\ \bibinfo {pages}
		{066101} (\bibinfo {year} {2010})}\BibitemShut {NoStop}%
	\bibitem [{\citenamefont {Sidorov}\ and\ \citenamefont
		{Sadykov}(2005)}]{Sidorov}%
	\BibitemOpen
	\bibfield  {author} {\bibinfo {author} {\bibfnamefont {V.}~\bibnamefont
			{Sidorov}}\ and\ \bibinfo {author} {\bibfnamefont {R.}~\bibnamefont
			{Sadykov}},\ }\href@noop {} {\bibfield  {journal} {\bibinfo  {journal}
			{Journal of Physics: Condensed Matter}\ }\textbf {\bibinfo {volume} {17}},\
		\bibinfo {pages} {S3005} (\bibinfo {year} {2005})}\BibitemShut {NoStop}%
	\bibitem [{\citenamefont {Mutti}\ \emph {et~al.}(2011)\citenamefont {Mutti},
		\citenamefont {Cecillon}, \citenamefont {Elaazzouzi}, \citenamefont {Goc},
		\citenamefont {Locatelli}, \citenamefont {Ortiz},\ and\ \citenamefont
		{Ratel}}]{Nomad}%
	\BibitemOpen
	\bibfield  {author} {\bibinfo {author} {\bibfnamefont {P.}~\bibnamefont
			{Mutti}}, \bibinfo {author} {\bibfnamefont {F.}~\bibnamefont {Cecillon}},
		\bibinfo {author} {\bibfnamefont {A.}~\bibnamefont {Elaazzouzi}}, \bibinfo
		{author} {\bibfnamefont {Y.~L.}\ \bibnamefont {Goc}}, \bibinfo {author}
		{\bibfnamefont {J.}~\bibnamefont {Locatelli}}, \bibinfo {author}
		{\bibfnamefont {H.}~\bibnamefont {Ortiz}}, \ and\ \bibinfo {author}
		{\bibfnamefont {J.}~\bibnamefont {Ratel}},\ }\href@noop {} {\bibfield
		{journal} {\bibinfo  {journal} {Proceedings of ICALEPCS2011}\ ,\ \bibinfo
			{pages} {808}} (\bibinfo {year} {2011})}\BibitemShut {NoStop}%
	\bibitem [{\citenamefont {Igarashi}\ \emph {et~al.}(2008)\citenamefont
		{Igarashi}, \citenamefont {Christensen}, \citenamefont {Larsen},
		\citenamefont {Olsen}, \citenamefont {Pedersen}, \citenamefont {Rasmussen},\
		and\ \citenamefont {Dyre}}]{Igarashi}%
	\BibitemOpen
	\bibfield  {author} {\bibinfo {author} {\bibfnamefont {B.}~\bibnamefont
			{Igarashi}}, \bibinfo {author} {\bibfnamefont {T.}~\bibnamefont
			{Christensen}}, \bibinfo {author} {\bibfnamefont {E.~H.}\ \bibnamefont
			{Larsen}}, \bibinfo {author} {\bibfnamefont {N.~B.}\ \bibnamefont {Olsen}},
		\bibinfo {author} {\bibfnamefont {I.~H.}\ \bibnamefont {Pedersen}}, \bibinfo
		{author} {\bibfnamefont {T.}~\bibnamefont {Rasmussen}}, \ and\ \bibinfo
		{author} {\bibfnamefont {J.~C.}\ \bibnamefont {Dyre}},\ }\href@noop {}
	{\bibfield  {journal} {\bibinfo  {journal} {Review of Scientific
				Instruments}\ }\textbf {\bibinfo {volume} {79}},\ \bibinfo {pages} {045106}
		(\bibinfo {year} {2008})}\BibitemShut {NoStop}%
	\bibitem [{\citenamefont {Casalini}\ and\ \citenamefont
		{Roland}(2003)}]{Casalini}%
	\BibitemOpen
	\bibfield  {author} {\bibinfo {author} {\bibfnamefont {R.}~\bibnamefont
			{Casalini}}\ and\ \bibinfo {author} {\bibfnamefont {C.~M.}\ \bibnamefont
			{Roland}},\ }\href {\doibase 10.1063/1.1624401} {\bibfield  {journal}
		{\bibinfo  {journal} {The Journal of Chemical Physics}\ }\textbf {\bibinfo
			{volume} {119}},\ \bibinfo {pages} {11951} (\bibinfo {year}
		{2003})}\BibitemShut {NoStop}%
	\bibitem [{\citenamefont {Grzybowska}\ \emph {et~al.}(2006)\citenamefont
		{Grzybowska}, \citenamefont {Pawlus}, \citenamefont {Mierzwa}, \citenamefont
		{Paluch},\ and\ \citenamefont {Ngai}}]{Grzybowska}%
	\BibitemOpen
	\bibfield  {author} {\bibinfo {author} {\bibfnamefont {K.}~\bibnamefont
			{Grzybowska}}, \bibinfo {author} {\bibfnamefont {S.}~\bibnamefont {Pawlus}},
		\bibinfo {author} {\bibfnamefont {M.}~\bibnamefont {Mierzwa}}, \bibinfo
		{author} {\bibfnamefont {M.}~\bibnamefont {Paluch}}, \ and\ \bibinfo {author}
		{\bibfnamefont {K.~L.}\ \bibnamefont {Ngai}},\ }\href {\doibase
		10.1063/1.2354492} {\bibfield  {journal} {\bibinfo  {journal} {The Journal of
				Chemical Physics}\ }\textbf {\bibinfo {volume} {125}},\ \bibinfo {pages}
		{144507} (\bibinfo {year} {2006})}\BibitemShut {NoStop}%
	\bibitem [{\citenamefont {Angell}\ \emph {et~al.}(2000)\citenamefont {Angell},
		\citenamefont {Ngai}, \citenamefont {McKenna}, \citenamefont {McMillan},\
		and\ \citenamefont {Martin}}]{Angell}%
	\BibitemOpen
	\bibfield  {author} {\bibinfo {author} {\bibfnamefont {C.~A.}\ \bibnamefont
			{Angell}}, \bibinfo {author} {\bibfnamefont {K.~L.}\ \bibnamefont {Ngai}},
		\bibinfo {author} {\bibfnamefont {G.~B.}\ \bibnamefont {McKenna}}, \bibinfo
		{author} {\bibfnamefont {P.~F.}\ \bibnamefont {McMillan}}, \ and\ \bibinfo
		{author} {\bibfnamefont {S.~W.}\ \bibnamefont {Martin}},\ }\href {\doibase
		10.1063/1.1286035} {\bibfield  {journal} {\bibinfo  {journal} {Journal of
				Applied Physics}\ }\textbf {\bibinfo {volume} {88}},\ \bibinfo {pages} {3113}
		(\bibinfo {year} {2000})}\BibitemShut {NoStop}%
	\bibitem [{\citenamefont {Berthier}\ and\ \citenamefont
		{Biroli}(2011)}]{Berthier}%
	\BibitemOpen
	\bibfield  {author} {\bibinfo {author} {\bibfnamefont {L.}~\bibnamefont
			{Berthier}}\ and\ \bibinfo {author} {\bibfnamefont {G.}~\bibnamefont
			{Biroli}},\ }\href@noop {} {\bibfield  {journal} {\bibinfo  {journal}
			{Reviews of Modern Physics}\ }\textbf {\bibinfo {volume} {83}},\ \bibinfo
		{pages} {587} (\bibinfo {year} {2011})}\BibitemShut {NoStop}%
\end{thebibliography}
\end{document}